\documentclass[fleqn,usenatbib]{mnras}

\usepackage{mathptmx}

\usepackage[T1]{fontenc}
\usepackage{ae,aecompl}

\usepackage{graphicx}	
\usepackage{amsmath}	
\usepackage{amssymb}	
\usepackage{lscape} 
\usepackage{url} 

\newcommand{\Lsun}{$\mathrm{L_{\sun}}$}
\newcommand{\Ledd}{$L_{\mathrm{Edd}}$}
\newcommand{\kms}{km\,s$^{-1}$}
\newcommand{\MBH}{$M_{\mathrm{BH}}$}
\newcommand{\civ}{\ion{C}{iv}}
\newcommand{\um}{$\mu$m}
\newcommand{\Li}{$L_{\mathrm{i}}$}

\title[FIR emission associated with nuclear outflows]{Far-infrared emission in luminous
quasars accompanied by nuclear outflows}

\author[N. Maddox et al.]{
Natasha Maddox$^{1}$\thanks{E-mail: maddox@astron.nl},
M.J. Jarvis$^{2,3}$, M. Banerji$^{4,5}$, P.C. Hewett$^{4}$,  
N. Bourne$^{6}$, 
\newauthor L. Dunne$^{6,7}$, S. Dye$^{8}$, S. Eales$^{7}$,
C. Furlanetto$^{9}$, S.J. Maddox$^{6,7}$,
\newauthor M.W.L. Smith$^{7}$, E. Valiante$^{7}$\\
$^{1}$ASTRON, the Netherlands Institute for Radio Astronomy, Postbus 2,
7990 AA, Dwingeloo, The Netherlands\\
$^{2}$Oxford Astrophysics, Denys Wilkinson Building,
University of Oxford, Keble Rd, Oxford, OX1 3RH, UK\\
$^{3}$Physics Department, University of the Western Cape,
Cape Town, 7535, Republic of South Africa\\
$^{4}$Institute of Astronomy, Madingley Road, Cambridge CB3 0HA, UK\\
$^{5}$Kavli Institute of Cosmology Cambridge, Madingley Road, Cambridge CB3 0HA, UK\\
$^{6}$Institute for Astronomy, University of Edinburgh, Royal Observatory, Edinburgh EH9 3HJ, UK\\
$^{7}$School of Physics and Astronomy, Cardiff University, The Parade, Cardiff, CF24 3AA, UK\\
$^{8}$School of Physics and Astronomy, University of Nottingham, University Park, Nottingham, NG7 2RD, UK\\
$^{9}$Instituto de F\'{i}sicia y Astronom\'{i}a, Universidad de Valpara\'{i}so, Avda. Gran Breta\~{n}a 1111, Valpara\'{i}so, Chile\\
}

\date{Accepted XXX. Received YYY; in original form ZZZ}

\pubyear{2017}

\begin{document}
\label{firstpage}
\pagerange{\pageref{firstpage}--\pageref{lastpage}}
\maketitle

\begin{abstract}

Combining large-area optical quasar surveys with the new
far-infrared Herschel-ATLAS Data Release 1, we search for an
observational signature associated with the minority of quasars
possessing bright far-infrared (FIR) luminosities.
We find that FIR-bright quasars show broad \civ\ emission line
blueshifts in excess of that expected from the optical luminosity
alone, indicating particularly powerful nuclear outflows. The quasars
show no signs of having redder optical colours than the general
ensemble of optically-selected quasars, ruling out differences in
line-of-sight dust within the host galaxies. We postulate that these
objects may be caught in a special 
evolutionary phase, with unobscured, high black hole accretion rates and 
correspondingly strong nuclear outflows. The high FIR emission found
in these objects is then either a result of star formation related to
the outflow, or is due to dust within the host galaxy 
illuminated by the quasar. We are thus directly witnessing coincident
small-scale nuclear processes and galaxy-wide activity, commonly
invoked in galaxy simulations which rely on feedback from quasars to
influence galaxy evolution.

\end{abstract}

\begin{keywords}
surveys--galaxies:evolution--galaxies:star formation--quasars:general--infrared:galaxies

\end{keywords}


\section{Introduction}\label{sec:Introduction}

The coincidence of the peak of star formation (SF) density
(\citealt{Hopkins2008}) and accretion in active galactic nuclei (AGN,
\citealt{Richards2006}) indicates a link between the two
processes. This connection is further reinforced by the relationship between
the mass of galaxy bulges and the central black holes they host
(\citealt{Magorrian1998}, \citealt{Ferrarese2000}). Some interaction between nuclear activity
and SF in the host galaxies, generally referred to as `feedback', is
invoked to facilitate this link. Simulations routinely insert feedback
from AGN to quench SF and prevent galaxies from becoming arbitrarily
large by heating and/or expelling the gaseous fuel for future SF
(\citealt{Bower2006}, \citealt{Sijacki2007}). The mechanism for expelling the gas is
generally considered to be an outflow, driven either by radiation, or
radio jets, originating with the AGN. 

Feedback from AGN on galaxy (\citealt{Morganti2005}) and even cluster
scales (\citealt{Rawlings2004}) is readily 
observed at radio and X-ray wavelengths, in a process commonly known
as `radio mode' or `kinetic' feedback (see \citealt{Fabian2012} for 
a review). On smaller scales, radio jets are seen to interact with the
intergalactic medium, clearing gas from the nuclear region
(e.g. \citealt{Morganti2013}). Radiation, or `quasar mode'
feedback from luminous quasars\footnote{We use the term `AGN' to
  describe an actively accreting black hole, and `quasar' to denote
  the luminous, unobscured subpopulation of AGN.}, is more difficult to observe
directly. Collections of objects at low redshift
(\citealt{Mullaney2013}, \citealt{Cicone2014}), and
high redshift (\citealt{Harrison2016}, \citealt{Brusa2015}) show
indications of outflowing gas in the properties of narrow emission
lines. Broad absorption line quasars
(BALQSOs) show clear evidence of rapid outflows driven by the AGN,
reaching speeds of many thousands of kilometres per second 
(\kms; \citealt{Weymann1981}). Similar outflows can also be observed in X-rays
(\citealt{Reeves2003}, \citealt{Reeves2009}). The energy injected, and
ultimately the effect on the BALQSO host 
galaxy, is difficult to determine, except under special circumstances
(\citealt{Moe2009}, \citealt{Dunn2010}, \citealt{Capellupo2014}). 

In the context of feedback, and the coincidence of the SF and AGN
histories, directly observing signatures of the interaction between the AGN and
galaxy-wide SF is complicated by the very small spatial scales
encompassing the black hole environment, and our inability to
spatially resolve these scales for all but the most local systems. An additional
complicating fact is that emission from luminous, unobscured
quasars dominates the spectral energy distribution (SED) at most
wavelengths, with the exception of the far-infrared (FIR;
  defined herein as 60--500\um), where 
dust heated by star formation is a significant emitter
(\citealt{Schweitzer2006}, \citealt{Hatz2010}). With the arrival of sensitive, relatively high
spatial resolution FIR imaging over the large areas required to
contain the rare, luminous quasars, we now have the data to probe
samples spanning a wide range of intrinsic luminosities and redshifts.

Recent investigations decomposing the SEDs of quasars show
contributions at FIR wavelengths both from warm dust, heated by the
AGN, and cooler dust, heated by star formation throughout the galaxy
(\citealt{Pece2015}, \citealt{Symeonidis2016}, \citealt{Symeonidis2017}). These
studies find the relative 
contributions and resulting FIR luminosity vary from object to object, with
some systems showing an excess of FIR flux, and others remaining
undetected at these long wavelengths. While optically more luminous
quasars are more likely to be detected in a FIR-flux limited sample, not all of the most
luminous quasars are FIR-detected. Conversely, optically less luminous
quasars, while less frequently FIR-bright, do also show FIR emission. 
Irrespective of a putative underlying correlation between FIR and
accretion luminosity, the observations show that for a given accretion
luminosity, a wide range of FIR luminosities are possible. 

Here we investigate the minority of optically selected quasars which are
observed to be FIR bright, searching for some observational signature
that can be linked with the cause of their FIR flux, incorporating data from the
\textit{Herschel}\footnote{\textit{Herschel} is 
  an ESA space observatory with science instruments provided by
  European-led Principal Investigator consortia and with important
  participation from NASA} Space Observatory
(\citealt{Pilbratt2010}). In Section~\ref{sec:data} we describe the
data used for the study, and investigate the quasar
properties in Section~\ref{sec:nuclearoutflows}. A discussion and
conclusions are presented in Section~\ref{sec:discussion}. Concordance
cosmology with $H_{0} = 70$ km s$^{-1}$ Mpc$^{-1}$ (thus $h\equiv
H_{0}$/[100 km s$^{-1}$ Mpc$^{-1}$]$=0.7$), $\Omega_{m} = 0.3$,
$\Omega_{\Lambda} = 0.7$ is assumed throughout.

\section{Optical quasars and FIR data}\label{sec:data}

We use FIR photometry from the largest-area survey
undertaken with the \textit{Herschel} Space Observatory, the
\textit{Herschel} Astrophysical 
Terahertz Large Area Survey (H-ATLAS, \citealt{Eales2010}). The full survey
covers 600 square degrees in five infrared bands centred at 100,
160 (PACS, \citealt{Poglitsch2010}), and 250, 350 and 500\um\ (SPIRE,
\citealt{Griffin2010}). We focus on the three equatorial fields centred on
RA = 9, 12, and 15h, contained within the Data Release 1 (DR1, 
\citealt{Valiante2016}). These three fields cover 161.6 deg$^2$, and have
extensive multiwavelength ancillary data, crossmatching for which is described in
\citet{Bourne2016}. The catalogues can be downloaded from the H-ATLAS
website\footnote{\url{http://www.h-atlas.org/public-data/download}}. 

The three equatorial fields have been covered by the Sloan Digital Sky
Survey (SDSS, \citealt{York2000}). We extract quasars from the Data
Release 7 (DR7, \citealt{Schneider2010}) and Data Release 10 quasar
catalogues (DR10, \citealt{Paris2014}). For objects common to both surveys,
the order of preference is SDSS DR10 $>$ SDSS DR7, as the DR10 spectra
have larger wavelength coverage and generally higher signal-to-noise
ratio (SNR). The different flux limits and selection algorithms for
DR7 and DR10 are complementary, and result in a combination of surveys with
essential dynamic range in luminosity and redshift, less affected by
selection effects known to plague individual surveys. The number of
objects from each survey contributing to the final quasar sample is listed in
Table~\ref{tab:quasars}. The full sample of 4938 quasars spans $0<z<5$,
and several orders of magnitude in optical luminosity. Objects 
associated with radio sources from the Faint Images of the Radio Sky at
Twenty-centimeters survey (FIRST, \citealt{Becker1995}) are excluded
from the sample, to avoid contamination of the FIR from synchrotron
emission. The final sample contains 4667 quasars.

We wish to explore beyond the nominal 4-$\sigma$ flux limit of the
published FIR catalogues, so we extract 250\um\ fluxes and 1-$\sigma$
uncertainties at the coordinates of the optical quasars directly from
the maps, downloaded from the H-ATLAS webpage. This is possible
because we know with high precision the positions of the quasars
within the maps. For the fluxes, we use the
\texttt{HATLAS\_GAMA$<$FIELD$>$\_DR1\_FILT\_BACKSUB$<$BAND$>$.FITS}
maps, where \texttt{FIELD} is 9, 12 or 15, and \texttt{BAND} is 250.
We set the mean of the maps to zero to remove any remaining offsets. The corresponding
noise is extracted from the associated noise maps,
\texttt{HATLAS\_GAMA$<$FIELD$>$\_DR1\_FILT\_SIGMA$<$BAND$>$.FITS}. The
noise maps only include instrumental noise, so confusion noise derived from
Equation 14 of \citet{Valiante2016} and a 5.5~per~cent calibration term
are added in quadrature to produce the final 1-$\sigma$ uncertainty.

The map astrometry is accurate enough to use the flux in the maps at the
6\arcsec\ pixel corresponding to the optical position of the
quasars. The flux densities are monochromatic, and assume $F_{\nu} \propto
\nu^{-1}$, where $F_{\nu}$ is the flux, and $\nu$ is frequency. High
redshift quasars are unresolved in the 18\arcsec\ 
\textit{Herschel} beam. Comparing the
250\um\ map fluxes and uncertainties for our extracted 
4-$\sigma$ sources with those from the published catalogues show good
agreement. 

We restrict our analysis to sources with 2-$\sigma$ 250\um\
detections, increasing the number of quasars with FIR fluxes from 352
at 4-$\sigma$ to 1185 at 2-$\sigma$. Our results are unchanged if we
use only 3- or 4-$\sigma$ sources, at the expense of poorer statistics.
As we deal here only with fluxes, we are not required to make any 
assumptions about the emission mechanism for the FIR radiation. Based
on SED decomposition of FIR-detected quasars (\citealt{Hatz2010},
\citealt{Pece2015}), the observed 250\um\ fluxes will have an
increasing contribution from warm dust in the torus heated by the AGN
with increasing redshift, combined with flux from cooler dust heated
by star formation. That some combination of AGN and SF-heated dust is
producing FIR flux is sufficient for this investigation.

\begin{table}
\centering
  \caption{Numbers of quasars from each survey contributing to the
    final sample. Subset is restricted to $1.6\le z\le 4.8$.}
\label{tab:quasars}
\begin{tabular}{lrrr} \hline
Survey & Number & 2-$\sigma$ & 4-$\sigma$ \\ \hline
DR10 & 3209 & 675 & 183 \\
DR7 & 1458 & 510 & 169 \\
Total & 4667 & 1185 & 352 \\ 
Subset &  2704 & 565 & 149 \\ \hline
\end{tabular}
\end{table}

\section{Evidence of nuclear outflows in FIR-bright
  quasars}\label{sec:nuclearoutflows}

Here we investigate the ultra-violet (UV) spectral properties of the FIR-detected
quasars to gain insight into their extreme FIR
luminosity. While high redshift quasars are
spatially unresolved, their spectroscopic details allow us to
probe material at a variety of galacto-centric radii, including the
innermost parsecs surrounding the AGN. We focus on the \civ\ emission line, as it
is a bright, prominent spectral feature that can be studied in
moderate resolution, moderate SNR spectra, and is found to show large variations in
its properties which correlate with a number of other quasar
observables (\citealt{Richards2011}).

\subsection{Measuring Redshifts and Blueshifts}\label{subsec:redblue}

The redshifts for both the SDSS DR7 and DR10 quasars have been
determined using a mean field independent component analysis (MFICA)
scheme, as described in \citet{Allen2013}. For each quasar the
spectrum is reconstructed using the MFICA components and the redshift
is determined simultaneously. Conceptually, the approach is identical
to that used to give the $z_{PCA}$ redshifts of the SDSS DR12 quasars
(\citealt{Paris2017}), where components derived from a principal
component analysis (PCA) are employed (e.g. \citealt{Francis1992};
\citealt{Paris2011}). The MFICA-derived redshifts show significantly
reduced systematic bias as a function of the form of the quasar spectra (Allen
\& Hewett 2017, in preparation). In the context of the investigation
presented in this paper, the redshift improvements are not critical
and all results would remain essentially unchanged if, instead,
redshifts for SDSS DR7 quasars from \citet{Hewett2010} and
$z_{PCA}$ for SDSS DR10 quasars (\citealt{Paris2017}) were employed.

To calculate the \civ\ blueshift we first define a power-law
continuum, $f(\lambda) \propto \lambda{^\alpha}$, with the slope,
$\alpha$, determined using the median values of the flux in two
continuum windows at 1445--1465 and 1700--1705\AA. The power-law
continuum is subtracted from the spectrum. The \civ\ emission
line is taken to extend over the wavelength interval 1500--1600\AA
(corresponding to approximately $\pm 10\,000$\,\kms\ from the
rest-frame transition wavelength), a recipe that is commonly adopted
(e.g. \citealt{Denney2013}). Narrow absorption features, which are
frequently found superimposed on \civ\ emission, are identified
and a simple interpolation scheme employed to `fill in' the emission
line flux at wavelengths affected by the absorption.  The blueshift of
the \civ\ emission line, in \kms, is defined as $c \times
(1549.48 - \lambda_{half})/1549.48$, where $c$ is the velocity,
1549.48\AA \ is the rest-wavelength of the \civ\ emission, assuming
equal contributions from both components of the doublet, and
$\lambda_{half}$ is the wavelength which bisects the cumulative
\civ\ emission line flux. This procedure has been found to be less
sensitive to problems arising from low SNR spectra. Full details of
the \civ\ measurements can be found in \citet{Coatman2016}.

Our redshift range is restricted to $1.6\le z\le 4.8$ to ensure 
coverage of the \civ\ feature in the SDSS spectra. BALQSOs have been
excluded, as accurate parametrization of the emission line and
blueward continuum are not possible for these objects.

\subsection{Blueshifted C\,{\sc{iv}} in FIR-bright quasars}\label{subsec:civ}

Blueshifted \civ\ emission lines in quasar spectra are nearly
ubiquitous, with the magnitude of the blueshift showing correlations
with a variety of other properties (see \citealt{Richards2011} for a
comprehensive investigation).  \civ\ blueshifts are associated with
disk-wind driven nuclear outflows (\citealt{Gaskell1982},
\citealt{Murray1995}, \citealt{Richards2011}), and thus provide
information about the conditions on small spatial scales, in the
immediate environment of the central supermassive black hole. Large \civ\
blueshifts are also associated with quasars accreting at high
Eddington ratios \citep{Coatman2016}. This phase is expected to 
be short-lived (\citealt{Hopkins2005}, \citealt{Hopkins2010}).

\begin{figure}
\includegraphics[width=\columnwidth]{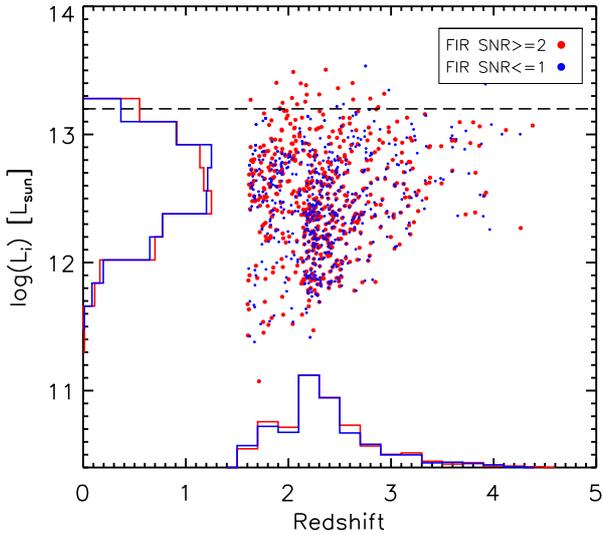}
\caption{The $i$-band luminosity and redshift ranges for the
  quasars detected in the FIR at $\ge$2-$\sigma$ (red points), and
  the matched sample of quasars undetected at $\le$1-$\sigma$ (blue
  points). The histograms on the bottom and side show the normalised
  distributions of the respective populations in redshift and \Li.} 
\label{fig:MatchedLi}
\end{figure}

We create matched FIR-detected and FIR-undetected samples in order to
compare the \civ\ blueshifts for the FIR-bright quasars with respect 
to the general quasar population. For every SNR $\ge 2$ FIR-detected
quasar, we choose a quasar with FIR SNR $\le 1$ at the closest
redshift and $i$-band magnitude. For cases where one FIR-undetected
object is the closest match to several FIR-detected objects, as is
common at bright luminosities where most objects are FIR-detected, the
next nearest neighbour is chosen if it is within a distance of 0.3 in
the redshift--Log(\Li) plane. If there is no suitable next nearest
neighbour, the duplicated closest match is accepted.
The ranges in $i$-band luminosity,
\Li, and redshift for the resulting matched samples is shown in
Fig.~\ref{fig:MatchedLi}. We restrict the sample to have log(\Li)$ < 13.2$
\Lsun\ due to a lack of FIR-undetected sources at the highest
luminosities. Constructing a matched sample removes the
known underlying correlation between \civ\ blueshift and quasar
luminosity (\citealt{Richards2011}), assuming the colours of the two
sub-samples are similar (but see Section~\ref{subsec:dust}). A
two-sided Kolmogorov-Smirnov (KS) test on the redshift and $i$-band 
magnitude distributions of the matched samples return probabilities
of 0.999863 and 0.984483, respectively, thus there is no evidence for
rejecting the null hypothesis that the two histograms are drawn from
the same underlying distributions of these two observables.

\begin{figure}
\includegraphics[width=\columnwidth]{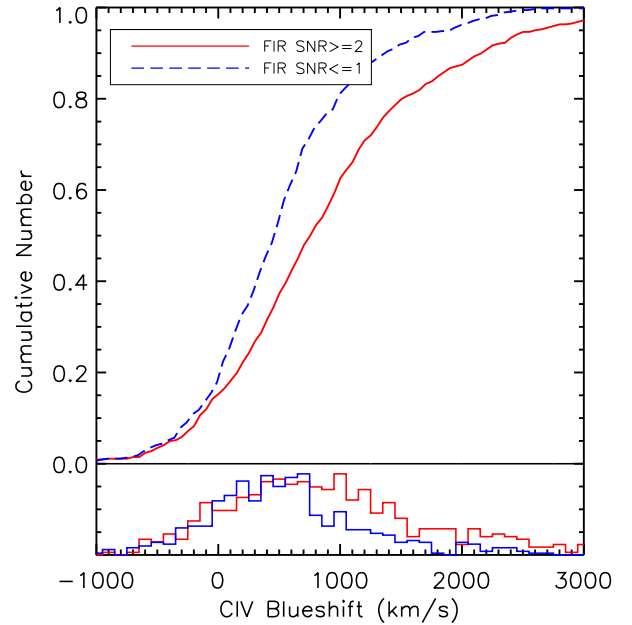}
\caption{Cumulative distribution of the blueshift of the \civ\
  emission line for the quasars detected in the FIR at
  $\ge$2-$\sigma$ (red solid line), and those undetected at
  $\le$1-$\sigma$ (blue dashed line). The FIR-detected quasars have
  larger \civ\ blueshifts, indicating more prominent nuclear-driven
  outflows. The bottom panel shows normalised histograms of the FIR-detected and
  FIR-undetected \civ\ blueshifts, in red and blue, respectively.}
\label{fig:CIVblueshift}
\end{figure}

Fig.~\ref{fig:CIVblueshift} shows the cumulative distributions of
\civ\ blueshift for the FIR-detected and FIR-undetected matched
samples. The FIR-detected quasars show larger
blueshifts than is expected from their FIR-undetected counterparts, with a
two-sided KS test returning a probability that the two
distributions are drawn from the same underlying sample of
10$^{-12}$. Using 3- and 4-$\sigma$ FIR detections instead of the
2-$\sigma$ detections, the KS test returns probabilities of 10$^{-8}$
and 0.0001 that the FIR-detected and FIR-undetected distributions are
from the same underlying samples. The 4-$\sigma$ sample only contains
140 FIR-detected objects, and thus suffers from the small sample size.
Recall that BALQSOs have been removed from the sample, so the
  difference can not be attributed to absorbed flux in the blue wing
  of the emission line. While a nuclear outflow is not always accompanied by a
FIR excess, for the sample as a whole, at a given optical luminosity,
FIR-bright quasars have larger \civ\ blueshifts than FIR-undetected quasars.

\subsection{Red or Reddened Quasars}\label{subsec:dust}

Vigorous star formation is known to be a dusty process, 
so if the FIR flux is at least partly from star formation, we may expect
the FIR-detected quasars to be reddened by dust in the 
host galaxies, as evidenced by redder optical colours with respect to
the general quasar population. Using the same
FIR-detected and undetected matched samples as before, we compare the
SDSS $g-r$ colours of the two samples. For these luminous quasars, the
optical photometry is entirely dominated by emission from the quasar, even if
accompanied by a starburst in the host galaxy. To remove the
redshift dependence of quasar colours, the colour of an unreddened model quasar
similar to that from \citet{Maddox2008} is subtracted from each object to create a
`normalised' colour. Using the normalised colour, with the redshift
dependency removed, creates a distribution which better illustrates
departures from the median colour of a quasar at any redshift. 

\begin{figure}
\includegraphics[width=\columnwidth]{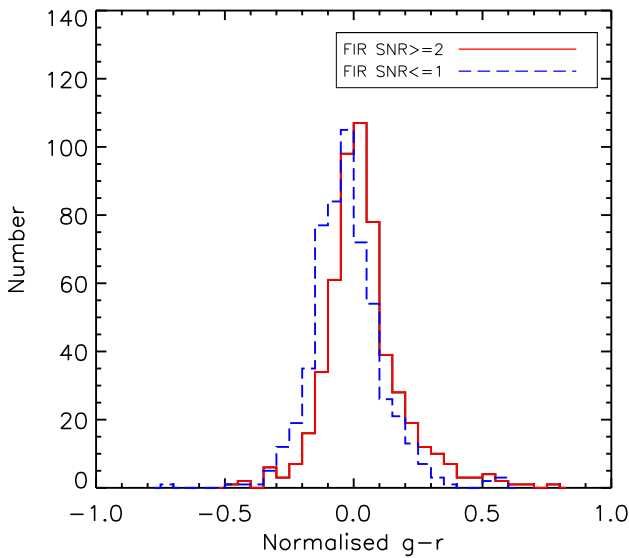}
\caption{$g-r$ colour for FIR and FIR non-detected quasars, normalised
by the $g-r$ colour for the unreddened model quasar as a function of redshift. The
FIR-detected quasars are redder than the non-detected quasars, but the
difference is due to the known different emission line properties of
the two sub-samples, not differences in the continuum, which would be
expected from dust in the host galaxies.}
\label{fig:gmr}
\end{figure}

The results are shown in Fig.~\ref{fig:gmr}. The FIR-detected quasars 
indeed appear to be redder than the FIR-undetected sample, 
with a KS-test showing the probability that the two
distributions are from the same underlying sample is
10$^{-14}$. However, upon closer examination, the bulk of the offset
arises from the known differences in the \civ\ emission line
properties between the two sub-samples. From \citet{Richards2011},
large \civ\ blueshifts are accompanied by smaller \civ\ equivalent
width. As the \civ\ emission line enters the SDSS $g$-band at $z\sim
2$, the $g-r$ colours for FIR-detected quasars are not as blue as those
for the FIR-undetected sub-sample. The difference in the $g-r$
distribution disappears when the different \civ\ emission line
properties are accounted for by using a version of the
\citet{Maddox2008} model with altered emission line strengths. The
KS-test probability goes from 10$^{-14}$ to 0.03, and we are no longer
able to say that the underlying distributions are significantly different.
The lack of obvious signs of excess dust obscuration in the
FIR-detected quasar host galaxies is consistent with the finding in
\citet{Richards2003} and \citet{Richards2011} that quasars with large
\civ\ blueshifts are less likely to have red continua.

This result does not rule out the possibility that a number of
FIR-bright quasars with moderate host galaxy dust content may be
missing from current quasar catalogues. Both the FIR-detected and
undetected quasars were selected at optical (or at these high
redshifts, rest-frame UV) wavelengths, which are sensitive to even
small amounts of dust. Although the SDSS quasar selection does include
some objects suffering from moderate dust extinction, these objects
are rare, and the vast majority of quasars have colours consistent
with very little dust obscuration. Selecting quasars at wavelengths
less sensitive to dust reddening, such as the NIR, would provide a
quasar sample less biased against obscuration.

\section{Discussion}\label{sec:discussion}

The FIR fluxes for quasars at a given redshift and optical luminosity 
range from $>$5-$\sigma$ detections to non-detections. Potential
sources of the exceptional emission in the FIR-bright 
quasars are dust heated from galaxy-wide star formation, dust heated
by the AGN, or most likely, some combination of the two. 
At low redshifts, the 250\um\ flux is a relatively uncontaminated
measure of the cool dust heated solely by star formation. However, at the redshifts
probed here, 250\um\ corresponds to rest-frame 83, 62 and 50\um\ for
objects at $z=2, 3, 4$, and SED modeling has shown that the 
contribution to these wavelengths from dust in the nuclear torus
heated by the AGN is substantial (\citealt{Hatz2010},
\citealt{Pece2015}). 

Dust within the host galaxy heated by the quasar itself is also a
possible source of the FIR emission. \citet{Symeonidis2016} make a case for
AGN-heated dust being the primary source of FIR emission at low
redshift via SED decomposition, and find the result holds for higher
redshift, more luminous sources (\citealt{Symeonidis2017}). The nuclear
outflows observed for the sample of objects presented here may facilitate
this heating by enabling the quasar radiation to reach further into
the host galaxy, and interacting with more dust.
Full SED decomposition is required to determine the relative
contributions of SF-heated dust and AGN-heated dust, as they may have 
different temperatures, which we defer to further work. 

A connection between AGN-driven nuclear outflows and galaxy-scale gas
content is consistent within the framework of AGN feedback. Simulations 
from \citet{Hopkins2010} show that it is indeed possible 
for an AGN outflow to affect gas in the quasar host galaxies at
appreciable (kiloparsec) distances from the nucleus. These simulations were
performed in the context of an evolutionary 
sequence, where a gas-rich major merger triggers a heavily
dust-obscured starburst, observed as a ULIRG-type object, accompanied
by obscured black hole accretion. The radiation pressure from AGN accretion is
then responsible for clearing out the gas and dust, quenching the star
formation via nuclear outflows, and revealing an unobscured, broad-line quasar.
The timescale for this is relatively short, on the order of 100Myr. 

We find that our FIR-bright quasars show no indication of redder
colours with respect to the FIR-faint population. However, as the
quasars were selected at optical--UV wavelengths, the sensitivity of
the quasar selection to objects suffering even small amounts of dust
reddening is low, thus there is the possibility of a dusty, FIR-bright quasar
population excluded from the current sample. Work
by \citet{Ishibashi2016} shows that a dusty interstellar medium
facilitates effective feedback by assisting 
a radiative wind to exert pressure on the medium and
ultimately expel it from the galaxy. Within this context, the
heavily dust-reddened objects uncovered in \citet{Banerji2012} and 
\citet{Banerji2015} would be extreme examples of these objects. 

Recent work by \citet{Coatman2016} and \citet{Coatman2017} show that
quasars with \civ\ blueshifts $>1000$\,\kms\ have overestimated black hole
masses (\MBH) when directly derived from the shape of the \civ\
emission line, by a factor of five for blueshifts of 3000\kms, and as
much as an order of magnitude in the most extreme 
cases. Therefore, for a given luminosity, the Eddington luminosity 
ratio, $L/$\Ledd, is substantially higher for the high blueshift
objects, resulting in stronger nuclear winds. From
Fig.~\ref{fig:CIVblueshift}, it is around 1000\,\kms\ 
where the strongest discrepancy between the blueshift values for the
FIR-detected and undetected samples diverges. Simulations predict that
high Eddington ratio accretion is relatively short-lived 
(\citealt{Hopkins2010}). In addition, since the \MBH\ is
under-massive, and active accretion indicates rapid black hole growth,
we may be witnessing an early, mass-building phase.

\citet{Coatman2016} also suggest that the high \civ\ blueshift quasars
is the parent population from which BALQSOs are drawn. Crossmatching
our full sample of 4667 quasars to the SDSS DR6 BALQSO catalogue of
\citet{Allen2011}, and the BAL classifications for SDSS DR12 quasars
from \citet{Paris2017}, we find 284 radio-quiet matches, of which 81
are 2-$\sigma$ FIR detections, or 28$\pm$3.2~per~cent. For a sample of
non-BALQSOs matched in redshift and $i$-band magnitude, the
FIR-detected fraction is 23$\pm$2.8~per~cent. This only includes
high-ionisation \civ\ BALQSOs (HiBALs), not low-ionization BALQSOs
(LoBALs) or FeLoBALs, all of which are known to have redder colours
than the general quasar population (\citealt{Reichard2003}). The
numbers are small, but it does indicate that BALQSOs are at least as likely to be FIR-luminous
as non-BALQSOs, and possibly even more likely, in contrast to
the result from \citet{Cao2012}. In special circumstances, the
distance to the absorbing material, and thus the mass of the material
and energy of the outflow, can be estimated, and it is found that the
energy contained within the outflow is sufficient to have an effect on
galaxy scales (\citealt{Moe2009}, \citealt{Capellupo2014}). The
outflows with distance estimates put the absorbing material at
kiloparsec-scale distances, far beyond the immediate black hole
neighbourhood, consistent with the radial extent found within simulations.


\section{Conclusions}\label{sec:conclusions}

From the large number of quasars contained within the 161~deg$^2$ of
DR1 H-ATLAS data, we are able to investigate the properties of
the sample while accounting for underlying and observational
trends. We find that, for a given optical luminosity, FIR-detected
quasars have stronger nuclear outflows, as measured by the blueshift
of the \civ\ emission line, than a luminosity and redshift-matched
FIR-undetected quasar sample. For the same luminosity and
redshift-matched samples, the FIR-detected quasars have redder optical
colours, but this difference is primarily due to differences in the
emission line properties of the two sub-samples. 

The temporal coincidence of nuclear outflow signatures, along with strong FIR
emission, result in a plausible scenario where we are
witnessing the phase where the central AGN is clearing out its
environment, and is accompanied either by the tail end
of the initial ULIRG-like starburst, or the subsequent triggered
starburst. Alternatively, the FIR emission could be from dust within
the host galaxies heated by the quasar itself.


\section*{Acknowledgements}

We thank the anonymous referee for helpful comments which improved this paper. 
NM wishes to thank R. Morganti, M. Michalowski, and G. De Zotti for
useful comments. MB acknowledges funding from the Science and
Technology Facilities Council via an Ernest Rutherford Fellowship. PCH
acknowledges support from the STFC via a Consolidated Grant to the
Institute of Astronomy, Cambridge. LD and SJM acknowledge funding from
the European Research Council Advanced Investigator grant, COSMICISM and
also from the ERC consolidator grant CosmicDust. The
\textit{Herschel}-ATLAS is a project with Herschel, which is an ESA space 
observatory with science instruments provided by European-led
Principal Investigator consortia and with important participation from
NASA. The H-ATLAS website is \url{http://www.h-atlas.org/}.

Funding for the SDSS and SDSS-II has been provided by the Alfred
P. Sloan Foundation, the Participating Institutions, the National
Science Foundation, the U.S. Department of Energy, the National
Aeronautics and Space Administration, the Japanese Monbukagakusho, the
Max Planck Society, and the Higher Education Funding Council for
England. The SDSS Web Site is \url{http://www.sdss.org/}.

Funding for SDSS-III has been provided by the Alfred P. Sloan
Foundation, the Participating Institutions, the National Science
Foundation, and the U.S. Department of Energy Office of Science. The
SDSS-III web site is \url{http://www.sdss3.org/}. 









\bsp	
\label{lastpage}
\end{document}